\documentclass[11pt]{article}
\def\be{\begin{equation}}
\def\ee{\end{equation}}
\def\ba{\begin{eqnarray}}
\def\ea{\end{eqnarray}}
\def\la{\langle}
\def\ra{\rangle}

\def\m{\mu}

\def\h{\hskip 1cm}

\usepackage{epsfig}
\begin{document}
\begin{titlepage}

\vspace{4cm}

\begin{center}{\Large \bf Exact solutions for a universal set of quantum gates on a family of iso-spectral spin chains
}\\
\vspace{1cm} V. Karimipour\footnote{Corresponding Author, email:
vahid@sharif.edu},
\hspace{0.5cm} N. Majd \footnote{email:naymajd@mehr.sharif.edu}\\
\vspace{1cm} Department of Physics, Sharif University of Technology,\\
P.O. Box 11365-9161,\\ Tehran, Iran
\end{center}
\vskip 3cm

\begin{abstract}
We find exact solutions for a universal set of quantum gates on a
scalable candidate for quantum computers, namely an array of two
level systems. The gates are constructed by a combination of
dynamical and geometrical (non-Abelian) phases. Previously these
gates have been constructed mostly on non-scalable systems and by
numerical searches among the loops in the manifold of control
parameters of the Hamiltonian.
\end{abstract}
\end{titlepage}

\section{Introduction}
 Let $H(t)=H(R(t))$ be a time dependent Hamiltonian acting on an
$n$ dimensional Hilbert space, whose dependence on time is via
some control parameters collectively denoted by $R(t)$. Let this
hamiltonian have a $k$-dimensional degenerate subspace $V_0(t)$
spanned by the instantaneous eigenvectors $|1(t)\ra, |2(t)\ra,
\cdots |k(t)\ra$, with energy $E_0(t)$. When the point $R(t) $
moves around a loop in the space of control parameters, any state
$|\psi(0)\ra \in V_0$ evolves into a state

\begin{equation}\label{U}
    |\psi(t)\ra \equiv U(t) |\psi(0)\ra = e^{i\beta(t)}\Gamma |\psi(0)\ra.
\end{equation}
Here
\begin{equation}\label{beta}
    e^{i\beta(t)}:=e^{-i \int_{0}^{t} {E_0(t')dt'}},
\end{equation}
is the dynamical phase where $E_0(t)$ is the instantaneous
eigenvalue of the subspace $V_0$ and
\begin{equation}\label{Gamma}
 \Gamma:=P(e^{\oint_C {\bf A(R)}\cdot dR})=T(e^{\oint_C {\bf A(t)}\cdot dt}),
\end{equation}
is the non-abelian geometric phase which is the holonomy operator
associated with the anti-hermitian connection $A$ given by
\begin{equation}\label{A}
    A_{\m; ij}(R) = \la i(R)|\frac{\partial}{\partial R^{\m}}|j(R)\ra,
\end{equation}
or
\begin{equation}
A_{ij}(t) \equiv  A_{\m; ij}(R)\frac{dR^{\m}}{dt}= \la
i(t)|\frac{d}{dt}|j(t)\ra.
\end{equation}
 Note that the symbols $P$ and $T$ in the first and the second
integral of (\ref{Gamma}) refer respectively to the path ordering
and time ordering of the exponential around the loop $C$ in the
control manifold. The basic property of a general holonomy operator
is that it is independent of the way the loop is traversed in the
parameter space. In some special cases it depends only on a few
basic geometrical properties of the loop, like its area. In our case
which will be discussed in detail later, in which the connection is
constant, the holonomy depends only on this constant connection
(which is nothing but the tangent vector on the loop at $t=0$), and
the total time needed for traversing the loop. Note that in the
above formulas, the parameter $t$ does not necessarily point to
time, although we use this word for explicitness. It points to
any single parameter which parameterizes the loop $C$.\\
 In the special case when $V_0$ is one
dimensional, $\Gamma$ is the abelian geometrical phase and
identical to the well known Berry
phase. \\
This general scheme when applied to the field of quantum
computation takes the name of holonomic or geometrical quantum
computation and the unitary operators thus obtained are called
holonomic quantum gates. The problem of exact holonomic
implementation of quantum gates is of great interest in the field
of quantum computation
\cite{zanras,preskill,ellinas,pachos,cen,zhang,ericsson,nak1,nak2,km}.
This is due to the fact that holonomic quantum computation, being
geometrical in nature has a degree of stability against a class of
errors
\cite{preskill, ellinas, kuzmin}.\\
In particular it is known that these gates depend only on the loop
and not on the speed with which they are traversed. Moreover this
stability is related to the robustness of such gates against small
perturbations of the traversed loops \cite{preskill, ellinas,
kuzmin}
and against various noises.\\
 In the past few years many theoretical
proposals for holonomic implementation of quantum gates have been
reported in the literature \cite{ jones, duan, ericsson} and some
of them have been realized experimentally \cite{jones}. At present
we can say that there have been only sporadic successes in
overcoming one or the other of the many obstacles in the way of a
successful implementation of holonomic gates.\\ Among these
problems the requirement of scalability is the most important one.
Let us briefly discuss this issue.\\ It is well known that any
unitary gate can be constructed to arbitrary precision from a
combination of a universal set of gates. There are many choices
for this universal set. One choice is the set $\{H, P(\phi),
C(\pi)\}$, where $H = \frac{1}{\sqrt{2}}\left(
\begin{array}{cc}
  1 & 1 \\
  1 & -1 \\
\end{array}
\right)$ is the Hadarmard gate, $P(\phi) = \left(
\begin{array}{cc}
  1 &0 \\
  0 & e^{i\phi} \\
\end{array}
\right)$ is the phase gate and $C(\phi)={\rm
diag}(1,1,1,e^{i\phi})$ is the controlled phase-gate. If we can
act by these universal set of gates on any single qubit or any two
qubits of our scalable system, then we can say that holonomic
quantum gates have been constructed on the scalable system.\\
We stress that such gates should be implemented on a scalable
system, which is a crucial requirement for any viable candidate for
quantum computation.  A scalable candidate of quantum computer takes
the form of an array of two level systems, or qubits. The array of
identical systems can of course have higher dimensions but only two
of their states will play the role of our computational qubit. Since
any operation between remote qubits can be divided into elementary
logic gates on adjacent qubits, it is sufficient to enact the
universal set of gates only on single qubits and two adjacent
qubits.\\

An essential property of a scalable system is that the two qubit
gates be realized on the tensor product of the same space on which
the single qubit gates have been realized. However most of the
proposals of holonomic computation so far suggested, lack this
property. For example it has been shown that by abelian holonomy or
Berry phase, one can implement the one-qubit phase gate on a single
spin subject to a time varying magnetic field, and the two-qubit
conditional phase gate on a pair of of coupled spins
\cite{ericsson}. On the other hand the Hadamard gate which is needed
to be added to the above set if we are to have a universal set of
gates, requires non-abelian holonomy \cite{zanras}, and is much more
difficult to realize than the other gates. In fact the proposal for
its holonomic realization is based on a completely different system
consisting
of two degenerate qubits and two ancilla qubits \cite{bergman, duan}. \\
In other words the hadamard gate is not implemented on the same
qubit on which the phase gate was implemented. \\

Of particular interest to us are the models based on iso-spectral
Hamiltonians which is reviewed in section (\ref{isosp}). Although
in this case one can easily calculate the time or path-ordered
exponential, the determination of exact solutions (exact loops in
the parameter space) which lead to a universal set of quantum
gates is difficult. Such an approach has been followed in
\cite{nak2}, but with a numerical search among the class of loops
for finding the required loop for each member of the universal
set. Moreover in the approach of \cite{nak2}, the single qubit
gates are constructed on two dimensional subspaces of a three
dimensional space, (i.e. $k=2, n=3$) and the two qubit gates are
constructed on four dimensional subspaces of a five dimensional
space, (i.e. $k=4,n=5$). This construction has the drawback that a
two-qubit gate is not constructed on the tensor product space of
two qubits, a requirement which is highly desirable for scalable
quantum computation. In a related work Niskanen , Nakahara and
Solomaa \cite{nak1} employ a three state hamiltonian of the form
\begin{equation}\label{hnak}
    {H_0}^{{\rm 1\ qubit}} \equiv \epsilon|2\ra\la 2| = \left(\begin{array}{ccc}
                  0 & 0 & 0 \\
                  0 & 0 & 0 \\
                  0 & 0 & \epsilon \\
                \end{array}
    \right)
\end{equation}
to implement the phase and the Hadamard gate on the degenerate
subspace $V_0={\rm Span} \{ |0\ra,|1\ra\}$ and the controlled
phase-gate on the degenerate subspace of the Hamiltonian
\begin{equation}\label{H2nak}
    {H_0}^{{\rm 2\ qubit}} =  {H_0}^{{\rm 1\ qubit}}\otimes I + I \otimes {H_0}^{{\rm
    1\ qubit}}.
\end{equation}
This construction can be generalized to $N-$ qubit case. The
dimension of the full Hilbert space scales as $3^N$.  The one and
the two qubit hamiltonians of the proposal of \cite{nak1} are
rather abstract and we do not know of any concrete realizations in terms of spins or some other suitable observables.\\

\textbf{Remark:} Perhaps it is not strictly correct to say that the
proposal of \cite{duan} is not scalable. In fact in this proposal
which implement the universal set $\{e^{i\theta \sigma_y}, P(\phi),
C(\pi)\}$, one can achieve scalability by going through polynomially
more steps and using ancilla bits which do not destroy the
scalability. The aim of this paper is to achieve scalability by
encoding each qubit in the two lowest states of two adjacent spins
in a spin chain as shown in figure (\ref{chain}). This is equivalent
to using s $N$ ancilla
qubits for the $N$ computational qubits. \\

What we want to show in this paper is that one can indeed find
exact solutions for universal gates on a scalable candidate for
qubits. For this aim, we consider a spin chain, and encode each
qubit into the Hilbert space of two adjacent spins. We then show
that by moving around appropriate loops in families of is-spectral
spin chains, (determined by the adjoint action of suitable
operators) one can implement a universal set of one and two qubit
gates on such a spin chain. Such gates are realized as a
combination of dynamical and geometrical non-abelian phases, when
the corresponding loops are traversed. For each gate there are a
set of parameters in the form of axis of rotations and frequencies
which when tuned suitably enact appropriate one and two qubit
gates on single and two adjacent qubits.

\begin{figure}[t]
\includegraphics[width=10cm,height=10cm,angle=0]{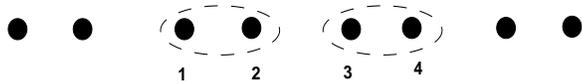}
\caption{In this array of spins, the degenerate ground state of
the two spins in a block, say spins 1 and 2 encode the two states
of a qubit. Exact holonomic gates act on each qubit and adjacent
qubits by suitable manipulations of spins.}
\end{figure}\label{chain}

The structure of this paper is as follows: In section
(\ref{isosp}) we review briefly the holonomies derived from
iso-spectral hamiltonians.
 In section (\ref{model}) we describe our model in detail where we show how a universal set of gates can be
 implemented by a combination of dynamical and geometrical phases on the spin chain. We end
 the paper with a discussion in section (\ref{dis}).\\

\section{Holonomies of iso-spectral Hamiltonians}\label{isosp}

In view of the time ordering of the exponential a closed formula
for the holonomy operator (\ref {A}) can not be obtained for most
connections
and one has to resort to numerical methods for an approximate calculation of this operator.\\
It would be much desirable if we could find connections whose
holonomy could be calculated exactly. This obstacle can be
partially overcome by restricting ourselves (and hence paying a
price of a using a limited source of holonomies) to constant
connections for which we have
\begin{equation}\label{Gammaconstant}
\Gamma = e^{AT},
\end{equation}
where $T$ is the total time required for traversing the loop.
\\
In view of the remark following equation (\ref{A}), the parameter
$T$ is not necessarily the total time needed for traversing the
loop, it is only the final value of the parameter $t$ which
parameterizes the loop, which in the sequel will be set to unity.
\\
Perhaps the simplest way for obtaining time-independent holonomy
operators is to consider iso-spectral family of Hamiltonians. These
are the hamiltonians which are of the form
\begin{equation}\label{iso}
    H(t):=e^{Xt}H_0e^{-Xt},
\end{equation}
where $X$ is any anti-hermitian operator and $H_0$ is the
hamiltonian at time $0$ with the degenerate subspace $V_0$,
spanned by the vectors $|1\ra, |2\ra, \cdots |k\ra$, i.e.
\begin{equation}\label{e0}
    H_0|i\ra = E_0|i\ra, \h i=1\cdots k.
\end{equation}
From relation (\ref{iso}) we find that at any time $t$, the
instantaneous eigenstates will have a simple form
\begin{equation}\label{et}
    |i(t)\ra = e^{Xt}|i\ra,\ \ \ H(t)|i(t)\ra = E_0|i(t)\ra, \ \ \
    i=1\cdots k.
\end{equation}

In this case we will find from (\ref{A}) that $A$ will be
constant, namely
\begin{equation}\label{xa}
    \la i|A|j\ra =  \la i|X|j\ra, \h i=1\cdots k.
\end{equation}
It is important to note that $X$ is an anti-hermitian operator
defined on the full Hilbert space and $A$ is an operator defined
only on the degenerate subspace and this relation implies only that
the projection of $X$ on this subspace is equal to $A$, that is
\begin{equation}\label{xx}
X|_{V_0} = A.
\end{equation}
Therefore a large number of operators $X$ can lead to the same
connection. Mathematically different loops in the parameter space
are specified by their tangent vector at the origin which is
nothing but the operator $X$.  \\
By re-scaling the time variable so that $T=1$, and taking into
account the inevitable dynamical phase, we arrive at the final
form of the operator $U$ acting on the space of a single qubit,
\begin{equation}\label{u}
    U=e^{-iE_0} e^{A}.
\end{equation}
\section{Holonomic computation on a spin chain}\label{model}

We take an array of qubits as shown in figure \ref{chain} so that
the spins within each block, say the spins (1,2) interact
according to the following Heisenberg hamiltonian:

\begin{equation}\label{1}
    H_0 = B({\sigma_{1z}}+ {\sigma_{2z}}) + J\overrightarrow{\sigma_1}\cdot \overrightarrow{\sigma_2},
\end{equation}
and different blocks do not interact with each other or their
interaction is so weak that we can consider them effectively
non-interacting at time $t=0$. Later on we will make these blocks
interact in order to implement two qubit gates. \
 It can be easily verified that if we choose the magnetic field so that $B=2J$, then the ground state of the
  Hamiltonian will
 be doubly degenerate.
 In fact the spectrum of the hamiltonian is as follows, where the states $|\phi_0\ra$ and $|\psi_0\ra$ are
 the degenerate ground states
 and $|\phi_1\ra$ and $|\phi_2\ra$ are the first and the second
 excited states.
\begin{eqnarray}
 \nonumber |\phi_2\ra &=& |+,+\ra , \h \h \h \ \ \h E = 5J,\\ \nonumber |\phi_1\ra &=& \frac{1}{\sqrt{2}}(|+,-\ra + |-,+\ra),
  \h \h E = J, \\ \nonumber
   |\phi_0\ra &=& |-,-\ra , \h \h \h \h \ \  E = -3J, \\  |\psi_0\ra &=& \frac{1}{\sqrt{2}}(|+,-\ra -
   |-,+\ra),\h \h \ \ E = -3J
  .
\end{eqnarray}
We take the code or computational qubits to be the degenerate
ground states, namely
\begin{equation}\label{qubits}
      |0\ra \equiv |\phi_0\ra=|-,-\ra, \h |1\ra \equiv |\psi_0\ra = \frac{1}{\sqrt{2}}(|+,-\ra-|-,+\ra).
\end{equation}
In spin notations where $S$ and $S_z$ respectively denote the
total and the $z$ component of the two spins, the code qubits are
$|0\ra = |S=1, S_z=-1\ra$ and $|1\ra = |S=0, S_z=0\ra $. We
suppose that these states are not hard to
access and control experimentally.\\
At low temperatures the two spins reside in the degenerate two
dimensional subspace which is a desirable
 situation for the initialization of the computer. If these two
 states were among the excited states of the Hamiltonian rather than the ground state, we would have been faced
 with an extra problem of exciting the two spins to these
 states. \\

Let the operator $X$ be of the following form
\begin{equation}\label{X}
    X =  i{{\bf n}}\cdot (\omega_1\overrightarrow{\sigma}_1 + \omega_2 \overrightarrow{
    \sigma}_2),
\end{equation}
which describes the rotation of the spins ${\bf S}_1$ and ${\bf
S}_2 $ around the axis ${{\bf n}}$ with frequencies $\omega_1$ and
$\omega_2$ respectively.\\
 From (\ref{A}) we find the gauge potential to be:

\begin{eqnarray}\label{gauge}
    A_{0,0} &=& \la \phi_0|X|\phi_0\ra = -i(\omega_1 + \omega_2)n_{z},\cr
    \nonumber A_{0,1} &=& \la \phi_0|X|\psi_0\ra =\frac{i}{\sqrt{2}}(\omega_1- \omega_2)(n_{x}+in_{y}),\cr
    \nonumber A_{1,0} &=& \la \psi_0|X|\phi_0\ra = \frac{i}{\sqrt{2}}(\omega_1- \omega_2)(n_{x}-in_{y}),\cr
   \nonumber A_{1,1} &=& \la \psi_0|X|\psi_0\ra=0.
\end{eqnarray}

Therefore in this subspace the gauge potential will be given by
the following operator
\begin{equation}\label{Asigma}
A =i(r_x \sigma_x + r_y \sigma_y + r_z\sigma_z+r_zI),
\end{equation}
where
\begin{equation}\label{rs}
  r_x = \frac{1}{\sqrt{2}}(\omega_1-\omega_2)n_{x}, \h
  r_y = \frac{-1}{\sqrt{2}}(\omega_1 -\omega_2)n_{y} ,\h
  r_z = \frac{-1}{2}(\omega_1 + \omega_{2})n_{z}.
  \end{equation}

After the lapse of time $T=1$ and acquiring the dynamical phase
$3J$, the gate
\begin{equation}\label{ufinal}
    u'=e^{i( r_x \sigma_x + r_y \sigma_y + r_z\sigma_z + (r_z+3J) I)},
\end{equation}
will act on the the space of single-qubit codes, $|0\ra$ and
$|1\ra$ . In this form the gate $u'$ is not general enough, since
the overall phase it applies namely $r_z+3J$ is not independent of
the other parameters. However at the end of any loop we can stop
changing the parameters of the Hamiltonian, and only pause for a
time interval $\tau$. This lapse of time will add a phase $3J\tau$
to the above phase and we obtain a general unitary gate given by
\begin{equation}\label{ufinall}
    u=e^{i( r_x \sigma_x + r_y \sigma_y + r_z\sigma_z + (r_z+3J(1+\tau)) I)},
\end{equation}

In this way by combining dynamical and geometrical phases we can
construct any single qubit gate on our code qubits, since the
parameters $r_x, r_y, r_z$ and $\tau$ are independent. It is only
necessary to choose the parameters $\omega_1, \omega_2, {\bf n} $
and $\tau $ appropriately. \\
We should emphasize that the specific structure of the degenerate
states, namely that it consists of a product state and an
entangled state, has been vital in our ability to arrive at a
general form of the holonomy $A\equiv X|_{V_0}$ with a simple
choice of the operator $X$. Had these two states been product
states, we should have used complicated and hence unjustified
forms of the
operator $X$ to arrive at the same general result. \\
Now let us explicitly construct the two single qubit gates in the
universal set, namely the phase gate $P(\phi)$ and the Hadamard
gate $H$.
\subsection{The phase gate}
The phase gate is defined as
\be
P(\phi)=\left(\begin{array}{cc}1 & 0 \\ 0 & e^{i\phi} \\
 \end{array}\right) = e^{i\frac{\phi}{2}(1-\sigma_z)+i2m\pi},
\ee where $m$ is an arbitrary integer.  Comparison with
(\ref{ufinall}) shows that for this gate we should have
\begin{equation}\label{rphase}
    r_x=r_y=0, \h r_z = -\frac{\phi}{2},\h
    r_z+3J(1+\tau)=\frac{\phi}{2} + 2m\pi.
\end{equation}

From (\ref{rs}) we find that the following choice of the
parameters implements this gate:
\begin{equation}\label{u1}
\hat{\bf n}=(0,0,1),\h \omega_1=\omega_2 = \frac{\phi}{2} ,\h
3J(1+\tau)= \phi+2m\pi.
\end{equation}
This is a rotation of both spins around the $z$ axis with equal
frequencies followed by a pause for a time interval $\tau$ given
as above. The freedom in choosing the integer $m$ guaranties that
the time lapse $\tau$ can always be positive as it should be.

\subsection{The Hadamard gate}
The Hadamard gate is \be H = \frac{1}{\sqrt{2}}\left(
\begin{array}{cc}
  1 & 1 \\
  1 & -1 \\
\end{array}
\right).\ee We now note that $H$ can be rewritten as follows:
\begin{equation}\label{Hlog}
    H = \frac{1}{\sqrt{2}}(\sigma_z + \sigma_x) =: {\bf
    k}\cdot \overrightarrow{\sigma},
\end{equation}
where ${\bf k}$ is a unit vector ${\bf k} =
\frac{1}{\sqrt{2}}(1,0,1)$. Since for any unit vector ${\bf k}$,
$e^{i\frac{\pi}{2}{\bf k}\cdot \overrightarrow{\sigma}}= i{\bf
k}\cdot \overrightarrow{\sigma}$, we find that
\begin{equation}\label{H}
H = -i \ i  {\bf k}\cdot \overrightarrow{\sigma} = -i
e^{i\frac{\pi}{2}{\bf k}\cdot \overrightarrow{\sigma}} =
e^{-i\frac{\pi}{2}} e^{i\frac{\pi}{2}{\bf k}\cdot
\overrightarrow{\sigma}}=e^{\frac{-i\pi}{2}I +
\frac{i\pi}{2\sqrt{2}}(\sigma_z+\sigma_x)+i2m\pi}.
\end{equation}

Comparison with (\ref{ufinal}) shows that the Hadamard gate is
produced when we choose the following parameters:
\begin{equation}\label{rH}
    r_x = r_z = \frac{\pi}{2\sqrt{2}}\h r_y = 0, \h r_z + 3J(1+\tau) =
    -\frac{\pi}{2}+ 2m\pi.
\end{equation}

Comparison with (\ref{rs}) determines the parameters of rotation
as follows:

\begin{equation}\label{HH}
    {\bf n} = (\sqrt{\frac{1}{3}}, 0 , -\sqrt{\frac{2}{3}}), \h
    \omega_1 = \frac{\pi}{2}\sqrt{3},\h \omega_2=0,\h
    3J(1+\tau)=-\frac{\pi}{2\sqrt{2}}(\sqrt{2}+1)+2m\pi.
\end{equation}

Thus we have constructed our single-qubit gates on our code space
which contains the computational qubits. We now turn to the
conditional phase gate to complete our universal set of gates.

\subsection{The conditional Phase gate}
The controlled phase gate has the following matrix form when the
basis vectors of the two qubits are ordered as $|0,0\ra,\ \
|0,1\ra,\ \  |1,0\ra\ $ and $\ |1,1\ra$:
\begin{equation}\label{Cphase}
    C(\phi)=\left(
\begin{array}{cccc}
  1 &  &  &  \\
   & 1 &  &  \\
   &  & 1 &  \\
   &  &  & e^{i\phi} \\
\end{array}\right) = exp \left(
\begin{array}{cccc}
  0 &  &  &  \\
   & 0 &  &  \\
   &  & 0 &  \\
   &  &  & i(\phi+2m\pi)\end{array}
\right).
\end{equation}

In view of the demanded scalability, this gate should act on the
space $V_0\otimes V_0$, where $V_0$ is the space on which the single
qubit gates act.  The space $V_0\otimes V_0$ is spanned by the
states of two qubits, namely
\begin{eqnarray}\label{2qubit}
|0,0\ra &=& |\phi_0\ra \otimes |\phi_0\ra = |-,-,-,-\ra,\cr
|0,1\ra &=& |\phi_0\ra \otimes |\psi_0\ra =
\frac{1}{\sqrt{2}}(|-,-,+,-\ra-|-,-,-,+\ra ),\cr|1,0\ra &=&
|\psi_0\ra \otimes |\phi_0\ra =
\frac{1}{\sqrt{2}}(|+,-,-,-\ra-|-,+,-,-\ra ),\cr|1,1\ra &=&
|\psi_0\ra \otimes |\psi_0\ra = \frac{1}{2}(|+,-,+,-\ra
-|+,-,-,+\ra-|-,+,+,-\ra+|-,+,-,+\ra) .
 \end{eqnarray}
To implement this gate the operator $X$ should act on the Hilbert
space of two adjacent pairs of spins, namely the pairs $(S_1, S_2)$
and $(S_3, S_4)$. Combining the geometric and the dynamical phases
we find that the gate which will be implemented on the two qubits is
equal to
\begin{equation}\label{u2}
    U^{{\rm{2 qubit}}} = e^{A+i(6J(1+\tau))},
\end{equation}

where  $A= X|_{V_0\otimes V_0}$  and we have used the fact that
the
energy of the degenerate subspace is now $6J$ instead of $3J$.\\
We now take the operator $X$ to be of the form

\begin{equation}\label{XX}
    X =  i\phi \left( \sigma_{2z}\sigma_{3z}+
    \sigma_{2z}+\sigma_{3z}\right).
\end{equation}
This operator couples the endpoint spins of the two neighboring
blocks which hitherto were considered non-interacting.

It is easy to verify that
\begin{eqnarray}
  X|0,0\ra &=& - i\phi |0,0\ra  \\ X|0,1\ra &=& - i\phi |0,1\ra \\
  X|1,0\ra &=& - i\phi |1,0\ra \\
  X|1,1\ra &=&  -i\phi |1,1\ra - 2i\phi |-,+,+,-\ra.
\end{eqnarray}
This will then lead to
\begin{equation}\label{XXX}
  A\equiv X|_{V_0\otimes V_0} =  \left(
\begin{array}{cccc}
  -i\phi &  &  &  \\
   & -i\phi &  &  \\
   &  & -i\phi &  \\
   &  &  & 0 \\
\end{array}
    \right).
\end{equation}
In view of (\ref{u2}) if we now choose $\tau$ so that $6J(1+\tau) =
\phi+2m\pi$, we will find
\begin{equation}\label{Af}
    A + i6J(1+\tau) = \left(
\begin{array}{cccc}
  0 &  &  &  \\
   & 0 &  &  \\
   &  & 0 &  \\
   &  &  & i(\phi+2m\pi) \\
\end{array}
    \right).
\end{equation}
and hence the conditional phase gate $C(\phi)$ will be exactly
implemented on the two qubits.\\ This completes our derivation of
exact holonomies for a universal set of gates on an array of
qubits.

\section{Discussion}\label{dis}

We have been able to implement a universal set of quantum gates on a
scalable system, by combining appropriately the dynamical and
non-abelian geometrical phases.  Our system consists of array of
two-spin blocks each of which is a four dimensional space with a two
dimensional degenerate subspace encoding the computational qubits.
With these universal set at hand, one can construct any other gate
to a sufficient degree of accuracy. The crucial step in this
direction has been an appropriate choice of a degenerate subspace of
a physical system which should represent the computational qubits.
This subspace should be so that the gauge connection projected on it
by a simple operator $X$ be general enough to represent an arbitrary
general gate. The choice of hamiltonian, so that its degenerate
subspace has one entangled state and one product state has been
essential in this step. Moreover we stress that the conditional
phase gate has been constructed on the tensor product of two such
qubits.  We should add that the experimental realization of such a
proposal is a completely different problem and we do not claim that
this proposal is superior to others as far as experimental
realization is concerned. We only emphasize the exact and the
scalable nature of the proposal and hope that following the basic
idea of this paper, namely taking two-spin blocks for representing
qubits, other researchers can proceed to more practical and
experimentally viable proposals.
\section{Acknowledgements}
We would like to thank A. Bayat, I. Marvian and D. Lashkari for
valuable discussions.

{}
\end{document}